\documentclass[%
aip,
jcp,
 sd,%
 amsmath,amssymb,
 reprint,%
]{revtex4-1}

\usepackage{graphicx}
\usepackage{dcolumn}
\usepackage{bm}
\usepackage{color}
\usepackage{ulem}

\begin{document}

\preprint{AIP/123-QED}

\title[Note: A replica liquid theory of binary mixtures]
{Note: A replica liquid theory of binary mixtures} 

\author{Harukuni Ikeda} \email{ikeda@r.phys.nagoya-u.ac.jp}
 \author{Kunimasa Miyazaki} \affiliation{ Department of Physics, Nagoya
 University, Nagoya, 464-8602, Japan }

\author{Atsushi Ikeda}
\email{atsushi.ikeda@phys.c.u-tokyo.ac.jp}
\affiliation{%
Graduate School of Arts and Sciences, University of Tokyo, 
Tokyo, 3-8-1, Japan
}%

\date{\today}

\maketitle


\newcommand{\diff}[2]{\frac{d#1}{d#2}}
\newcommand{\pdiff}[2]{\frac{\partial #1}{\partial #2}}
\newcommand{\fdiff}[2]{\frac{\delta #1}{\delta #2}}
\newcommand{\bx}{\bm{x}}
\newcommand{\by}{\bm{y}}
\newcommand{\new}{\nonumber\\}
\newcommand{\abs}[1]{\left|#1\right|}
\newcommand{\tr}{{\rm Tr}}
\newcommand{\obx}{\overline{\bm{x}}}
\newcommand{\oby}{\overline{\bm{y}}}
\newcommand{\onu}{\overline{\nu}}
\newcommand{\rhoog}{\rho_{\onu}^{\rm MG}(\obx)}
\newcommand{\rhodg}{\rho_{\onu}(\obx)}
\newcommand{\ox}{\overline{x}}
\newcommand{\oy}{\overline{y}}

The replica liquid theory (RLT) is a mean-field thermodynamic theory of
the glass transition of supercooled liquids\cite{parisi2010}. The theory
was first developed for one-component monatomic systems.  The RLT
enables one to predict the ideal glass transition temperature, $T_K$,
from a first-principles calculation, by considering the $m$ replicas of
the original system~\cite{parisi2010}.  Thermodynamic properties near
$T_K$ are deduced by computing the free energy of a liquid consisting of
$m$-atomic replica molecules and then taking the limit of $m\rightarrow
1$ at the end of the calculation.  The RLT was later extended to binary
systems~\cite{coluzzi1999,biazzo2009}.  However, it has been known that
the binary RLT is {\it inconsistent} with its one-component counterpart;
In the limit that all atoms are identical, the configurational entropy,
$S_c$, and thus $T_K$ calculated by the binary RLT differ from those
obtained by the one-component RLT~\cite{coluzzi1999,biazzo2009}.  More
specifically, an extra composition-dependent term, or the mixing
entropy, remains finite in $S_c$ computed by the binary RLT.  As
discussed by Coluzzi {\it et. al.}~\cite{coluzzi1999}, this
contradiction originates from the assumption that each replica molecule
consists of $m$-atoms of the {\it same} species.  Physically, this is
tantamount to assume that a permutation of atoms of one species with
atoms of the other species in a given glass configuration is
forbidden~\cite{coluzzi1999}.  This is indeed the case if, say, the
atomic radii of the two species are very different. Clearly this
assumption is inappropriate if the two species are very similar or
exactly identical because a permutation of the atoms of different
species are allowed.

In this Short Note, we reformulate the RLT in order to resolve this
problem.  We consider a binary liquid composed of $A$ and $B$ atoms.
 The important step is to rewrite the expression
of the grand canonical partition function in a form discussed by Morita
and Hiroike~\cite{morita1961} as
\begin{align}
 Z &= \sum_{N=0}^{\infty}\frac{1}{N!} \prod_{i=1}^N \sum_{\nu_i\in \{A,B\}} \int d\bx_i
\exp\left[-\beta V_N+\beta \sum_{i=1}^N\mu_{\nu_i}\right], \label{our}
\end{align}
where $\beta$ is the inverse temperature, $N$ is the total number of
atoms, $V_N$ is the total potential energy.  $\bm{x}_i$, $\nu_i\in \{A,
B\}$, and $\mu_{\nu_i}$ are the position, species, and chemical
potential of $i$-th atoms, respectively.  Eq.~(\ref{our}) is
mathematically equivalent to the standard expression for
$Z$~\cite{hansen1990}.  This expression can be readily generalized to
the replicated liquid consisting of $m$-atomic replica molecules as
\begin{align}
 Z_m &
= \sum_{N=0}^{\infty}\frac{1}{N!}\prod_{i=1}^N
 \left(
 \prod_{a=1}^m \sum_{\nu_{i}^a\in \{A,B\}}\int d\bx_i^a\right)\new
&\times \exp \left[-\beta \sum_{a=1}^m V_N^a + \beta \sum_{i=1}^N\psi_{\onu_i}(\obx_i) 
\right],
\label{eq2}
\end{align}
where $\psi_{\onu_i}(\obx_i) \equiv
\displaystyle{\sum_{a=1}^m\mu_{\nu_i^a} - u_{\onu_i}(\obx_i)}$ is the
generalized chemical potential in which the external potential
$u_{\onu_i}(\obx_i)$ is included.  $\overline{\bm{x}}_i \equiv
(\bm{x}_i^1,\cdots,\bm{x}_i^m)$ and $\overline{\nu}_i\equiv
(\nu_i^1,\cdots,\nu_i^m)$ denote the set of the positions and components
of $m$ atoms of the $i$-th molecule.  The advantage to express $Z_m$
\`{a} la Morita-Hiroike as Eq.~(\ref{eq2}) is that assigning a label of
the component $\overline{\nu}_i$ to each atom enables one to describe
replica molecules consisting of different set of species.
The density field conjugated to $\psi_{\onu}(\ox)$ can
be written as 
\begin{align}
 \rho_{\onu}(\obx) &= \sum_{i=1}^N\left\langle
\prod_{a=1}^m\delta(\bx^a-\bx_i^a)\delta_{\nu_i^a,\nu_i^a}
 \right\rangle  = \fdiff{\log Z_m}{\psi_{\onu}(\obx)}.\label{density}
\end{align}
Following the standard procedure~\cite{hansen1990}, one can express the
free energy, $-\beta F$, by the Legendre transformation from
$\psi_{\onu}(\obx)$ to $\rho_{\onu}(\obx)$, which can be written as the
sum of the ideal and the excess parts;
\begin{align}
\beta F[\rho_{\onu}(\obx)] &=
 -\sum_{\onu}\int d\obx \rho_{\onu}(\obx)(1-\log\rho_{\onu}(\obx))
+
\beta F_{\mbox{\scriptsize ex}}[\rho_{\onu}(\obx)].
\label{free2}
\end{align}

The equilibrium free energy is obtained by minimizing Eq.~(\ref{free2})
with respect to the density profile $\rho_{\onu}(\obx)$.  For
one-component systems, the standard procedure is to assume that
$\rho_{\onu}(\obx)$ is Gaussian-shaped and use its width, or the cage
size, as the minimization parameter~\cite{parisi2010}.  Once the
equilibrium free energy is obtained, $S_c$ is calculated by $S_c =
\lim_{m \to 1}m^2\pdiff{}{m} \frac{\beta F}{mN}$.  The ideal glass
transition temperature, $T_K$, is identified as the point at which
$S_c$ vanishes\cite{parisi2010}.  For binary liquids, however, the full
computation is a challenging task because the cage sizes vary
depending on the components $\onu$. But, at least, one can demonstrate
that the one-component result is correctly derived in the limit that
atoms of two components are identical or very similar.  In this limit,
the density profile in the replica space can be written as
\begin{align}
 \rho_{\onu}(\obx) = \rho \int d\bm{X}\prod_{a=1}^m \left(\sum_{\mu}
 c_{\mu}\delta_{\mu\nu^a}\gamma_{\Delta_\mu}(\bx^a-\bm{X})\right),\label{dg}
\end{align}
where $\gamma_A(\bm{x}-\bm{X})=e^{-|\bm{x}-\bm{X}|^2/2\Delta}/(4\pi
\Delta)^{d/2}$ is the Gaussian function centered at a reference position
$\bm{X}$ with the cage size $\sqrt{\Delta}$. $\rho=N/V$ is the number
density, and $c_\nu = N_\nu/N$ is the number fraction of the $\nu\in\{A,B\}$
species.  Eq.~(\ref{dg}) expresses that atoms of different species
constitute a single replica molecules with the composition ratio of
$c_A:c_B$.  This ansatz corresponds to the limit where a permutation
of the atoms of different species are allowed in a given glass
configuration.
The difference of the free energy from that of the
one-component system $F_{\rm 1}$ is expressed by the difference in the
ideal gas part (since the excess parts are identical in this limit) as
\begin{align}
\beta \Delta F &\equiv \beta \left\{F[\rho_{\onu}]- F_{\rm 1}[\rho]\right\}\new
& = -\sum_{\onu}\int d\obx \rhodg (1- \log \rhodg) \new
&+\int d\obx\rho(\obx)(1-\log\rho(\obx)),\label{154802_19Apr16}
\end{align}
where $\rho(\obx)$ is the density profile of the one-component system.
Because $\rho_{\onu}(\obx) = \rho(\obx)\times (\prod_a c_{\nu^a})$ in
the one-component limit, we arrive at $\beta \Delta F = mN \sum_{\mu} c_\mu
\log c_\mu$. This implies that the $S_c$ of the binary RLT correctly
converges to that of the one-component $S_{c,1}$, because
$\Delta S_c
\equiv S_c - S_{c,1}
 = m^2\partial_{m}\left({\beta \Delta F}/{mN}\right)  = 0.$

In the opposite limit where the atoms of the two species are very
different and one replica molecule consists of atoms solely of the one
species, one can show that the previous results of binary
RLT\cite{biazzo2009} are recovered.  In this case, the density profile
should be written as
\begin{align}
 \rho_{\onu}^{\prime}(\obx) &= \rho \int d\bm{X}\sum_{\mu} c_{\mu}\left(\prod_{a=1}^m
\delta_{\mu\nu^a}\gamma_{\Delta_\mu}(\bx^a-\bm{X})\right). \label{og}
\end{align}
Note the difference from Eq.~(\ref{dg}); The order of the product over
the atoms $a$ and the summation over the species $\mu$ have been
exchanged.  Due to the factor $(\prod_a \delta_{\mu\nu^a})$ in
Eq.~(\ref{og}), $\rho_{\onu}^{\prime}(\obx)$ vanishes unless each
molecule consists of a single species.  The non-vanishing component of
the density profile is $c_{\nu} \rho \int d\bm{X} \prod_a
\gamma_{\Delta_{\nu}}(\bx^a-\bm{X})$ ($\nu$= $A$ or $B$), which is
exactly the density field employed in previous studies of the binary
RLT~\cite{biazzo2009}. Eq.~(\ref{og}) should not be used in the
one-component limit because it gives a solution which is less stable
than Eq.~(\ref{dg}): Substituting Eq.~(\ref{og}) into Eq.~(\ref{free2})
and optimizing the parameter $\Delta_{\nu}$, one finds $\beta \Delta F =
N \sum_{\mu}c_\mu \log c_\mu$, which is larger than $\beta \Delta F$
from Eq.~(\ref{dg})\footnote{ Exchange ``larger'' with ``smaller'' for
$m<1$, see Ref.~\cite{parisi2010}.}.  This solution also leads to a
pathological result $\Delta S_c = -\sum_{\mu}c_\mu\log c_\mu > 0$; that
is, $S_c$ calculated assuming Eq.~(\ref{og}) is larger than the correct
one-component configurational entropy by the mixing
entropy\cite{ozawa2016}.

In summary, we reformulate the RLT of binary, or multi-component
mixtures, which correctly accounts for a permutation of the atoms in a
glass configuration and show that it resolves the inconsistency between
the one-component RLT and the binary RLT.  The binary RLT in the
previous studies is valid only in the limit where the atoms of different
species are so different that a permutation of the atoms of different
components is forbidden.  In the one-component limit, one has to
consider all possible permutations of atoms and adopts the density
profile expressed as Eq.~(\ref{dg}) to obtain the correct
configurational entropy. For general cases between
these two extreme limits, the density profile should be determined so as
to minimize the free energy, Eq.~(\ref{free2}).  Its implementation,
however, may be challenging.  
 The results discussed above imply a
possibility, for example, that a binary liquid of small and large hard
spheres undergoes the glass-glass transition from the {\it
``disordered'' glass} where the both types of spheres constitute a
replica molecule to the {\it ``ordered'' glass} where one replica
molecule consists solely of one of the component, as the size ratio
between the two components is varied, much the same way as the metallic
alloys undergo the order-disorder phase transition as the interactions
are varied~\cite{greiner2012}.  
 Note that this possibility has also been pointed out in
Ref.~\cite{ozawa2016} in the context of the numerical simulation of the
multi-component mixtures. Finally, we note that the
generalization to $n$-component system is straightforward by allowing
$\nu_i$ in Eq.~(\ref{our}) to take $n$-different states.  Polydisperse
case can be obtained in the $n\to\infty$ limit with a caveat about a
subtlety to take the continuous limit (see Ref.~[8])


\acknowledgments We are grateful to H. Yoshino, K. Hukushima, M. Ozawa, L. Berthier and F. Zamponi
for helpful discussions.  H. I. and K. M acknowledge 
JSPS KAKENHI Grants
Number JP16H04034, 
JP25103005, 
JP25000002, 
H. I. was supported by Program for Leading Graduate Schools
``Integrative Graduate Education and Research in Green Natural
Sciences'', MEXT, Japan.


%


\end{document}